\def\Rb{{I\!\! R}}
\def\Cb{\ \hbox{\vrule width 0.6pt height 6.5pt depth 0pt
              \hskip -3.2 pt} C}
\documentclass[aps,pra,showpacs,showkeys,preprint]{revtex4}
\usepackage{amsmath}
\usepackage{amssymb}
\usepackage{graphicx}
\begin{document}

\title{Generalized reduction criterion for separability
of quantum states}

\author{Sergio Albeverio$^{a}$\footnote{\emph{E-mail:}
albeverio@uni-bonn.de, SFB 611; IZKS; BiBoS; CERFIM(Locarno); Acc.
Arch.; USI(Mendriso)}}
\author{Kai Chen$^{b}$\footnote{\emph{E-mail:}
kchen@aphy.iphy.ac.cn}}
\author{Shao-Ming Fei$^{a,c}$\footnote{\emph{E-mail:} fei@uni-bonn.de}}

\affiliation{$~^{a}$ Institut f\"ur Angewandte Mathematik, Universit\"at Bonn, D-53115,
Germany\\
$~^{b}$ Laboratory of Optical Physics, Institute of Physics, Chinese Academy
of Sciences, Beijing 100080, P.R. China\\
$~^{c}$ Department of Mathematics, Capital Normal University, Beijing
100037, China}

\begin{abstract}
A new necessary separability criterion that relates the structures
of the total density matrix and its reductions is given. The
method used is based on the realignment method [K. Chen and L.A.
Wu, Quant. Inf. Comput. \textbf{3}, 193 (2003)]. The new
separability criterion naturally generalizes the reduction
separability criterion introduced independently in previous work
of [M. Horodecki and P. Horodecki, Phys. Rev. A \textbf{59}, 4206
(1999)] and [N.J. Cerf, C. Adami and R.M. Gingrich, Phys. Rev. A
\textbf{60}, 898 (1999)]. In special cases, it recovers the
previous reduction criterion and the recent generalized partial
transposition criterion [K. Chen and L.A. Wu, Phys. Lett. A
\textbf{306}, 14 (2002)]. The criterion involves only simple
matrix manipulations and can
therefore be easily applied.
\end{abstract}

\date{\today}
\maketitle

\section{Introduction}

In the last decade quantum entangled states have showed remarkable
applications and become one of the key resources in the rapidly
expanding field of quantum information processing. The history can
be traced back to the earlier well-known papers of Einstein,
Podolsky and Rosen \cite{EPR35}, Schr\"{o}dinger \cite{Sch35} and
Bell \cite{Bell64}. Recently quantum teleportation, quantum
cryptography, quantum dense coding and parallel computation
\cite{pre98,nielsen,zeilinger} have spurred a flurry of activity
in the effort to fully exploit the potential of quantum
entanglement. Despite of its importance, we do not yet have a full
understanding of the physical character and mathematical structure
for entangled states. We even do not know completely
wether a generic quantum state is entangled, and how much
entanglement remained after some noisy quantum processes.

A state of a composite quantum system is said to be \emph{disentangled}
or \emph{separable} if it can be prepared in a ``\emph{local}"
or \textquotedblleft \emph{classical}" way. A separable bipartite system
can be prepared as an ensemble realization of pure product states $%
\left\vert \psi _{i}\right\rangle _{A}\left\vert \phi _{i}\right\rangle _{B}$
($i=1,...,M$ for some positive integer $M$) occurring with a certain probability $p_{i}$:
\begin{equation}
\rho _{AB}=\sum_{i}p_{i}\rho _{i}^{A}\otimes \rho _{i}^{B},  \label{sep}
\end{equation}%
where $\rho _{i}^{A}=\left\vert \psi _{i}\right\rangle _{A}\left\langle \psi
_{i}\right\vert $, $\rho _{i}^{B}=\left\vert \phi _{i}\right\rangle
_{B}\left\langle \phi _{i}\right\vert $, $\sum_{i}p_{i}=1$ and $\left\vert
\psi _{i}\right\rangle _{A}$, $\left\vert \phi _{i}\right\rangle _{B}$ are
normalized pure states of the subsystems $A$ and $B$, respectively \cite%
{werner89}. If no convex linear combination exists for a given $\rho _{AB}$,
the state is called \textquotedblleft \emph{entangled}" and includes
quantum correlation.

For a pure state $\rho _{AB}$, it is straightforward to judge its
separability: \noindent a pure state $\rho _{AB}$ is separable if and only
if there is only one item in Eq.~(\ref{sep}) and $\rho ^{A}$ resp. $\rho ^{B}$ are the
reduced density matrices defined as $\rho ^{A}=Tr_{B}(\rho _{AB})$ and $\rho
^{B}=Tr_{A}(\rho _{AB})$. \noindent However, for a generic mixed state $\rho
_{AB}$, finding a decomposition like in Eq.~(\ref{sep}) or proving that it
does not exist is a non-trivial task (we refer to recent good reviews \cite%
{lbck00,terhal01,3hreview} and references therein). There has been
considerable effort in recent years to analyze the separability and
quantitative character of quantum entanglement. The Bell inequalities
satisfied by a separable system give the first necessary condition for
separability \cite{Bell64}. Many years after the appearance of Bell inequalities,
Peres made an important step
forward in 1996 by showing that partial transpositions with respect to one
and more subsystems of the density matrix for a separable state are
positive,
\begin{equation}
\rho ^{T_{\mathcal{X}}}\geq 0,
\end{equation}%
where $\mathcal{X}$ is either $A$ or $B$, $\rho ^{T_{\mathcal{X}}}$ stands
for the partial transpose with respect to $\mathcal{X}$. Thus $\rho ^{T_{\mathcal{X}}}$
should have non-negative eigenvalues (this is known as the $PPT$ criterion
or \emph{Peres-Horodecki} criterion) \cite{peres}, which was further
shown by Horodecki \textit{et al.} \cite{3hPLA223} to be sufficient for
$2\times 2$ and $2\times 3$ bipartite systems. Meanwhile, these authors also found
a necessary and sufficient condition for separability by establishing a
close connection between positive map theory and separability \cite{3hPLA223}%
. In view of the quantitative character for entanglement, Wootters succeeded
in computing the \textquotedblleft \textit{entanglement of formation}" \cite%
{be96} and thus obtained a separability criterion for $2\times 2$ mixtures
\cite{wo98}. The \textquotedblleft \textit{reduction criterion}" proposed
independently in \cite{2hPRA99} and \cite{cag99} gives another necessary
criterion which is equivalent to the $PPT$ criterion for $2\times n$
composite systems but is generally weaker. Pittenger \textit{et al.} gave
also a sufficient criterion for separability connected with the Fourier
representations of density matrices \cite{Rubin00}. Later, Nielsen \textit{%
et al.} \cite{nielson01} presented another necessary criterion called the
\textit{majorization criterion}: the decreasingly ordered vector of the
eigenvalues for $\rho _{AB}$ is majorized by that of $\rho _{A}$ or $\rho
_{B}$ alone for a separable state. A new method of detecting entanglement
called \textit{entanglement witnesses} was given in \cite{3hPLA223} and
\cite{ter00,lkch00}. Some necessary and sufficient criteria
of separability for low rank cases of the density matrix are also known
\cite{hlvc00,afg01,feipla02}. In addition, it was shown in \cite{wu00} and
\cite{pxchen01} that a necessary and sufficient separability criterion is also
equivalent to certain sets of equations. A \emph{PPT} extension based on
semidefinite programs is proposed in \cite{dps} which can test
numerically the separability.

However, despite these advances, practical and easily computable
criteria for a generic bipartite system are mainly limited to several ones such
as the $PPT$, reduction, majorization criteria as well as the \emph{PPT}
extension. Very recently Rudolph \cite{ru02} and K. Chen and L.A. Wu
\cite{ChenQIC03} proposed a new operational criterion for separability: the
\emph{realignment} criterion (named following the suggestion of \cite{Horo02},
it coincides with the \emph{computational cross norm} criterion given in
Ref. \cite{ru02}). The criterion is very simple to apply and shows dramatic
ability to detect many of the bound entangled states \cite{ChenQIC03} and even
genuinely tripartite entanglement \cite{Horo02}. Soon after the appearance
of \cite{ChenQIC03}, Horodecki
\textit{et al.} showed that the $PPT$ criterion and realignment criterion
are equivalent to the permutations of the density matrix's indices
\cite{Horo02}. A simple single framework for these criteria for the multipartite
case in any dimensions was recently given in \cite{chenPLA02} and is called the
\emph{generalized partial transposition criterion} (\emph{GPT}) which
includes, as special cases, the \emph{Peres-Horodecki} criterion
(\emph{PPT}), the \emph{realignment} criterion and the permutation indices
criterion for density matrix. Some further properties of the
\emph{realignment} criterion have been very recently derived in \cite{ru0212047}.

In this paper we present a systematic generalization of the reduction
criterion employing a realignment technique of a certain matrix constructed
from the density matrix. This criterion includes the
reduction criterion and the \emph{GPT} criterion as special cases. It
unifies them in a single simple framework. Thus our criterion is a strong separability
test for a generic bipartite or even for multipartite quantum states in
arbitrary dimensions. The detailed constructions are given in Section
\ref{sec2} where the reduction and the \emph{GPT} criteria are shown to be two special
cases of our new criterion. Some interesting examples are
given in Section \ref{sec3}. A brief summary and some discussions are
given in the last section.

\section{The criteria for separability}\label{sec2}

In this section we study the separability of the
density matrix for any bipartite system in arbitrary finite
dimension. Motivated by the
reduction criterion and the Kronecker product approximation technique \cite%
{loan,pits}, we give a new separability criterion by analyzing the trace norm
for some realigned version of a matrix constructed from the whole density
matrix and its reduced ones.

\subsection{Some notation}

\label{sec2.1} We first introduce some notations used in various matrix operations
(see e.g., \cite{hornt1,hornt}):

\noindent \textbf{Definition:} \emph{For any $m\times n$ matrix $A=[a_{ij}]$,
with entries $a_{ij}$, we define a vector $\mathcal{V}_{vec}(A)$ by}
\begin{equation*}
\mathcal{V}_{vec}(A)=[a_{11},\cdots ,a_{m1},a_{12},\cdots ,a_{m2},\cdots ,a_{1n},\cdots
,a_{mn}]^{t},
\end{equation*}
where $t$ represents the standard transposition operation.
Let $\mathcal{T}_{r}$ (resp. $\mathcal{T}_{c}$) denote the
row transposition (resp. column transposition) of $A$:
\begin{align}
\mathcal{T}_{r}(A)&=(\mathcal{V}_{vec}(A))^{t}, \label{p1} \\
\mathcal{T}_{c}(A)&=\mathcal{V}_{vec}(A).  \label{p2}
\end{align}
It is easily verified that
\begin{equation}\label{p3}
\mathcal{T}_{c}\mathcal{T}_{r}(A)=\mathcal{T}_{r}\mathcal{T}_{c}(A)=A^{t}.
\end{equation}
For example, for a $2\times 2$ matrix $A$,
$$
A=\left(
\begin{array}{cc}
a_{11} & a_{12} \\
a_{21} & a_{22}
\end{array}
\right),
$$
we have
\begin{align}
\mathcal{T}_{r}(A)&
=\left(\begin{array}{cc|cc}
a_{11} & a_{21} & a_{12} & a_{22}
\end{array}
\right) , \\
\mathcal{T}_{c}(A)&=\left(
\begin{array}{c}
a_{11} \\
a_{21} \\ \hline
a_{12} \\
a_{22}
\end{array}
\right) .
\end{align}

For a generic matrix $A={\displaystyle\sum\limits_{i,j}}
A_{ij}\left\vert i\right\rangle \left\langle j\right\vert ={\displaystyle
\sum\limits_{i,j}}A_{ij}\left\vert i\right\rangle \otimes \left\langle
j\right\vert ={\displaystyle\sum\limits_{i,j}}A_{ij}\left\langle
j\right\vert \otimes \left\vert i\right\rangle $, where $\left\vert
i\right\rangle ,\left\vert j\right\rangle $ are vectors of a suitably selected normalized
orthogonal basis, $\left\langle i\right\vert ,\left\langle j\right\vert $
are the corresponding transpositions (not conjugate transpositions).
Under the operations $\mathcal{T}_{r}$ and $\mathcal{T}_{c}$ one has:
\begin{align}
A\overset{\mathcal{T}_{r}}{\longrightarrow }{\displaystyle\sum\limits_{i,j}}%
A_{ij}\left\langle j\right\vert \otimes \left\langle i\right\vert \overset{%
\mathcal{T}_{c}}{\longrightarrow }{\displaystyle\sum\limits_{i,j}}%
A_{ij}\left\vert j\right\rangle \otimes \left\langle i\right\vert & =A^{t},
\label{p11} \\
A\overset{\mathcal{T}_{c}}{\longrightarrow }{\displaystyle\sum\limits_{i,j}}%
A_{ij}\left\vert j\right\rangle \otimes \left\vert i\right\rangle \overset{%
\mathcal{T}_{r}}{\longrightarrow }{\displaystyle\sum\limits_{i,j}}%
A_{ij}\left\vert j\right\rangle \otimes \left\langle i\right\vert & =A^{t}.
\label{p33}
\end{align}%
We further define $\mathcal{T}_{r_{k}}$ (resp. $\mathcal{T}_{c_{k}}$) ($k=A,B$)
to be the row (resp. column) transposition with respect to the subsystems $A,B$.
We set $\mathcal{T}_{\{x_{1},x_{2},...\}}\equiv\mathcal{T}_{x_{1}}\mathcal{T}_{x_{2}}...$
for $x_{1},~x_{2}\in \{r_{A},c_{A},r_{B},c_{B}\}$.
A generalized partial transposition operation (\emph{GPT}
operation) for a bipartite density matrix is thus given by $\mathcal{T}_{\mathcal{%
Y}}$, $\mathcal{Y}\subset \{r_{A},c_{A},r_{B},c_{B}\}$,
where $\mathcal{T}_{\mathcal{Y}}$ stands for all partial transpositions contained in
$\mathcal{Y}$ which is a subset of $\{r_{A},c_{A},r_{B},c_{B}\}$. With these
notations, the realignment criterion can
easily be stated. For example, for a bipartite system, we only need to make
partial transpositions with respect to $\mathcal{Y}=\{c_{A},r_{B}\}$.
This is equivalent to the realignment operation given in \cite%
{ru02} and \cite{ChenQIC03}, for the proof, see \cite{chenPLA02}.

We also need the following result in matrix analysis. \noindent
Let $Z$ be an $m\times m$ block matrix with block size $n\times
n$. We define a realigned matrix $\widehat{Z}$ of size
$m^{2}\times n^{2}$ that contains the same elements as in $Z$ but
in different positions,
\begin{equation}
\widehat{Z}=\left[
\begin{array}
[c]{c}
\mathcal{V}_{vec}(Z_{1,1})^{t}\\
\vdots\\
\mathcal{V}_{vec}(Z_{m,1})^{t}\\
\vdots\\
\mathcal{V}_{vec}(Z_{1,m})^{t}\\
\vdots\\
\mathcal{V}_{vec}(Z_{m,m})^{t}
\end{array}
\right]  .
\label{trans}
\end{equation}
$\widehat{Z}$ has the singular value decompositions:
\begin{equation}
\widehat{Z}=U\Sigma V^{\dagger }=\sum_{i=1}^{q}\sigma
_{i}u_{i}v_{i}^{\dagger },  \label{svd}
\end{equation}%
where $U=[u_{1}u_{2}\cdots u_{m^{2}}]\in \Cb^{m^{2}\times m^{2}}$
and $V=[v_{1}v_{2}\cdots v_{n^{2}}]\in \Cb^{n^{2}\times n^{2}}$
are unitary matrices, $\Sigma $ is a diagonal matrix with elements $\sigma _{1}\geq
\sigma _{2}\geq \cdots \geq \sigma _{q}\geq 0$ and $q\equiv\min (m^{2},n^{2})$.
In fact, the number of nonzero singular values $\sigma _{i}$ is the rank $r$
of the matrix $\widehat{Z}$, and $\sigma _{i}$ are exactly the nonnegative
square roots of the eigenvalues of $\widehat{Z}\widehat{Z}^{\dagger }$
or $\widehat{Z}^{\dagger }\widehat{Z}$ \cite{hornt}. Based on the above
constructions, $Z$ can be expressed as:
\begin{equation}
Z=\sum_{i=1}^{r}(X_{i}\otimes Y_{i}),  \label{sum}
\end{equation}
with $\mathcal{V}_{vec}(X_{i})=\sqrt{\sigma _{i}}u_{i}$ and $\mathcal{V}_{vec}(Y_{i})=\sqrt{\sigma _{i}}
v_{i}^{\ast }$ \cite{loan,pits}.

\subsection{The generalized reduction criterion for separability}
\label{sec2.2}

We now derive the
main result of this paper: a generalized reduction criterion for
separability of bipartite quantum systems in arbitrary dimensions.

\subsubsection{The main theorem}

The reduction criterion given in \cite{2hPRA99} and \cite{cag99} says that
for any bipartite $m\times n$ separable states, the following inequalities
should be satisfied simultaneously:%
\begin{equation}
\mathbb{I}_{m}\otimes \rho _{B}-\rho _{AB}\geqslant 0,\text{ \ \ \ \ \ }\rho
_{A}\otimes \mathbb{I}_{n}-\rho _{AB}\geqslant 0,  \label{reduction}
\end{equation}%
where $\rho _{A,B}$ are the reduced density matrixes with respect to
the subsystems $A$ and $B$, $\mathbb{I}_{m}$
(resp. $\mathbb{I}_{n}$) is an $m$ (resp. $n$) dimensional identity matrix.
This criterion is shown to be equivalent
to the \emph{PPT} criterion for $2\times n$ system but it is generally
weaker than the \emph{PPT} criterion \cite{2hPRA99,cag99}. So it certainly
cannot detect any bound entangled states which are \emph{PPT}. Noticing this
fact and the powerful ability of the realignment criterion, in particular
its generalization: the \emph{GPT} criterion, we expect that some stronger test
may appear. The reduction criterion is in essence a positive map.
Combining the technique for this positive map and the Kronecker product
approximation technique for a matrix \cite{loan,pits}, we apply a more
general map:
\begin{equation}\label{map}
\rho _{AB}\longrightarrow
\widetilde{\rho _{AB}}=ab\mathbb{I}_{mn}-a\mathbb{I}_{m}\otimes \rho
_{B}-b\rho _{A}\otimes \mathbb{I}_{n}+\rho _{AB},
\end{equation}
where $a,~b$ are arbitrary complex numbers. We are going to
derive a necessary separability condition of $\rho _{AB}$
in terms of the trace norm (Ky Fan norm) of a matrix obtained
from a $GPT$ map on $\widetilde{\rho _{AB}}$.
The trace norm of a matrix $Z$ is a unitary invariant
norm which is defined as the sum of all the singular values
of $Z$, or alternatively the sum of the nonnegative
square roots of the eigenvalues of $ZZ^{\dagger}$
or $Z^{\dagger}Z$. We thus arrive at the following separability
criterion for a bipartite system:

\vskip0.2cm \noindent \textbf{Theorem:} \emph{If a bipartite
density matrix $\rho _{AB}$ defined on an $m\times n$ space
is separable, then the generalized reduction
version $\widetilde{\rho _{AB}}$ of $\rho _{AB}$ should satisfy
\begin{equation}
||\widetilde{\rho _{AB}}^{\mathcal{T}_{\mathcal{Y}}}||\leq h_{a}h_{b},
~~~~~~\forall \mathcal{Y}\subset \{r_{A},c_{A},r_{B},c_{B}\},
\label{maintheorem}
\end{equation}
where $\mathcal{T}_{r_{k}}$ or $\mathcal{T}_{c_{k}}$ ($k=A,B$) stands for
transpositions with respect to the row or column for the subsystem $k$. The numbers
$h_{a},h_{b}$ are defined by
$$
\begin{array}{l}
\begin{array}{l}
h_{a}\equiv\\
\left\{
\begin{array}{ll}
|a-1|+(m-1)|a|, &r_{A},c_{A}\in \mathcal{Y}\text{\ or }~
r_{A},c_{A}\notin \mathcal{Y}, \\
(|a-1|^{2}+(m-1)|a|^{2})^{\frac{1}{2}},&r_{A}\in \mathcal{Y},~
c_{A}\notin \mathcal{Y}\text{\ or}\\
&c_{A}\in \mathcal{Y},~r_{A}\notin \mathcal{Y},
\end{array}
\right.
\end{array}\\
\begin{array}{l}
h_{b}\equiv\\
\left\{
\begin{array}{ll}
|b-1|+(n-1)|b|,&r_{B},c_{B}\in \mathcal{Y}\text{\ or }~
r_{B},c_{B}\notin \mathcal{Y}, \\
(|b-1|^{2}+(n-1)|b|^{2})^{\frac{1}{2}}, &r_{B}\in \mathcal{Y},~
c_{B}\notin \mathcal{Y}\text{\ or }\\
&c_{B}\in \mathcal{Y},~r_{B}\notin \mathcal{Y},
\end{array}
\right.
\end{array}
\end{array}
$$
where $a,~b\in\Cb$, $\mathcal{T}_{\mathcal{Y}}$ represents partial
transpositions with respect to every element contained in the set $\mathcal{Y}$ of the
related subsystems.}

\vskip0.2cm \noindent \textbf{Proof: }We only need to find the bound for the
trace norm of $||\widetilde{\rho _{AB}}^{\mathcal{T}_{\mathcal{Y}}}||$ with
respect to some operations \emph{$\mathcal{T}_{\mathcal{Y}}$} for any
separable states. Considering a separable bipartite system, we suppose $\rho
_{AB}$ has a decomposition, $\rho _{AB}=\sum_{i}p_{i}\rho _{i}^{A}\otimes
\rho _{i}^{B}$ with $0\leq p_{i}\leq 1$, $\sum_{i}p_{i}=1$. Under map (\ref{map})
it is evident that
$$
\begin{array}{rcl}
\rho _{AB}\rightarrow \widetilde{\rho _{AB}}
&=&ab\mathbb{I}_{mn}-a\mathbb{I}_{m}\otimes \rho _{B}-b\rho _{A}\otimes
\mathbb{I}_{n}+\rho _{AB}\\
&=&\sum_{i}p_{i}(a\mathbb{I}_{m}-\rho
_{i}^{A})\otimes (b\mathbb{I}_{n}-\rho _{i}^{B}),
\end{array}
$$
where $\rho _{A}$ and $\rho _{B}$ are the reduced density matrixes defined by $\rho
_{A}\equiv Tr_{B}(\rho _{AB})=\sum_{i}p_{i}\rho _{i}^{A}$, $\rho _{B}
\equiv Tr_{A}(\rho_{AB})=\sum_{i}p_{i}\rho _{i}^{B}$. We have
\begin{eqnarray*}
&&\rho _{i}^{A}\otimes \rho _{i}^{B} \\
&=&U_{i}\left(
\begin{array}{cccc}
1 &  &  &  \\
& 0 &  &  \\
&  & \ddots &  \\
&  &  & 0
\end{array}
\right) _{m}U_{i}^{\dag }\otimes V_{i}\left(
\begin{array}{cccc}
1 &  &  &  \\
& 0 &  &  \\
&  & \ddots &  \\
&  &  & 0
\end{array}
\right) _{n}V_{i}^{\dag },
\end{eqnarray*}
where we have diagonalized the (rank one) density matrix $\rho _{i}^{A}$
(resp. $\rho _{i}^{B})$
with the $m$ (resp. $n$)-dimensional unitary matrix $U_{i}$ (resp. $V_{i}$).
It is straightforward to check that
\begin{equation}
(a\mathbb{I}_{m}-\rho _{i}^{A})\otimes (b\mathbb{I}_{n}-\rho _{i}^{B})
=U_{i}\mathcal{A}U_{i}^{\dag }\otimes V_{i}\mathcal{B}V_{i}^{\dag },
\end{equation}
where $\mathcal{A}$ and $\mathcal{B}$ are diagonal matrices:
\begin{equation*}
\mathcal{A}=\left(
\begin{array}{cccc}
a-1 &  &  &  \\
& a &  &  \\
&  & \ddots &  \\
&  &  & a
\end{array}
\right) _{m},~\mathcal{B}=\left(
\begin{array}{cccc}
b-1 &  &  &  \\
& b &  &  \\
&  & \ddots &  \\
&  &  & b
\end{array}
\right) _{n}.
\end{equation*}
For an $m\times m$ (resp. $n\times n$) matrix $P$ (resp. $Q$) acting on the
complex space associated to the subsystem $A$ (resp. $B$),
the trace norm of the matrices $P$ and $Q$ has the property:
$||P\otimes Q||=||P||\cdot ||Q||$.
And for any matrices $X,~Y~Z$ acting on the
subsystem $A$ (or $B$), the $\mathcal{V}_{vec}$ operations
have the property: $\mathcal{V}_{vec}(XYZ)=(Z^{T}\otimes X)\mathcal{V}_{vec}(Y)$,
where both sides are column vectors and the tensor operation $\otimes$
has nothing to do with different subspaces $A$ and $B$ \cite{hornt}.

Let $\widetilde{\rho _{AB}}^{\mathcal{T}_{\mathcal{Y}}}$
denote the transformed matrix of $\widetilde{\rho _{AB}}$
under the partial transposition ${\mathcal{T}_{\mathcal{Y}}}$.
Without loss of generality
we suppose that we only make a row transposition to the subsystem $A$.
According to (\ref{p1}), we have
\begin{eqnarray}
&&(U_{i}\mathcal{A}U_{i}^{\dag }\otimes V_{i}\mathcal{B}V_{i}^{\dag })^{%
\mathcal{T}_{\{r_{A}\}}}  \notag \\
&=&(\mathcal{V}_{vec}(U_{i}\mathcal{A}U_{i}^{\dag }))^{t}\otimes V_{i}\mathcal{B}%
V_{i}^{\dag }  \notag \\
&=&(\mathcal{V}_{vec}(\mathcal{A}))^{t}(U_{i}{}^{\dag }\otimes U_{i}^{t})\otimes V_{i}%
\mathcal{B}V_{i}^{\dag }.
\end{eqnarray}%
In terms of  the unitary invariant property of the trace norm
we obtain
\begin{eqnarray*}
&&||(U_{i}\mathcal{A}U_{i}^{\dag }\otimes V_{i}\mathcal{B}V_{i}^{\dag })^{%
\mathcal{T}_{\{r_{A}\}}}|| \\
&=&||(\mathcal{V}_{vec}(\mathcal{A}))^{t}(U_{i}{}^{\dag }\otimes U_{i}^{t})||\cdot ||V_{i}%
\mathcal{B}V_{i}^{\dag }|| \\
&=&||(\mathcal{V}_{vec}(\mathcal{A}))^{t}||\cdot ||\mathcal{B}|| \\
&=&(|a-1|^{2}+(m-1)|a|^{2})^{\frac{1}{2}}(|b-1|+(n-1)|b|).
\end{eqnarray*}
Using the convex property of the trace norm, we get
\begin{eqnarray}
&&||\widetilde{\rho _{AB}}^{\mathcal{T}_{\{r_{A}\}}}||  \notag \\
&=&||\sum_{i}p_{i}(U_{i}\mathcal{A}U_{i}^{\dag }\otimes V_{i}\mathcal{B}%
V_{i}^{\dag })^{\mathcal{T}_{\{r_{A}\}}}||  \notag \\
&\leq &(\sum_{i}p_{i})(|a-1|^{2}+(m-1)|a|^{2})^{\frac{1}{2}}(|b-1|+(n-1)|b|)
\notag \\
&=&(|a-1|^{2}+(m-1)|a|^{2})^{\frac{1}{2}}(|b-1|+(n-1)|b|).
\end{eqnarray}%
A corresponding procedure can be applied for the column transposition to
the subsystem $A$ and corresponding operations for the subsystem $B$:
\begin{eqnarray*}
&&||\widetilde{\rho _{AB}}^{\mathcal{T}_{\{c_{A}\}}}|| \\
&\leq &(|a-1|^{2}+(m-1)|a|^{2})^{\frac{1}{2}}\cdot (|b-1|+(n-1)|b|), \\
&&||\widetilde{\rho _{AB}}^{\mathcal{T}_{\{r_{B}\}}}|| \\
&\leq &(|b-1|^{2}+(n-1)|b|^{2})^{\frac{1}{2}}\cdot (|a-1|+(m-1)|a|), \\
&&||\widetilde{\rho _{AB}}^{\mathcal{T}_{\{c_{B}\}}}|| \\
&\leq &(|b-1|^{2}+(n-1)|b|^{2})^{\frac{1}{2}}\cdot (|a-1|+(m-1)|a|).
\end{eqnarray*}

\noindent For the operations with respect to $\mathcal{Y}=\{r_{A},c_{A}\},$
we have in fact the \emph{PPT} operation to the subsystem $A$ of
$\widetilde{\rho _{AB}}$ and thus
\begin{eqnarray}
&&||\widetilde{\rho _{AB}}^{\mathcal{T}_{\{r_{A},c_{A}\}}}||  \notag \\
&&=||\sum_{i}p_{i}(U_{i}\mathcal{A}U_{i}^{\dag })^{t}\otimes V_{i}\mathcal{B}%
V_{i}^{\dag }||  \notag \\
&&\leq \sum_{i}p_{i}||U_{i}^{\ast }\mathcal{A}U_{i}^{t}\otimes V_{i}\mathcal{%
B}V_{i}^{\dag }||  \notag \\
&&=\sum_{i}p_{i}||\mathcal{A}\otimes \mathcal{B}||  \notag \\
&&=(|a-1|+(m-1)|a|)(|b-1|+(n-1)|b|).
\end{eqnarray}%
If the subsystem $A$ is left unchanged, i.e., both $c_{A},r_{A}\notin \mathcal{Y}$,
we have
\begin{eqnarray}
&&||\widetilde{\rho _{AB}}||  \notag \\
&&=||\sum_{i}p_{i}U_{i}\mathcal{A}U_{i}^{\dag }\otimes V_{i}\mathcal{B}%
V_{i}^{\dag }||  \notag \\
&&\leq \sum_{i}p_{i}||\mathcal{A}\otimes \mathcal{B}||  \notag \\
&&=(|a-1|+(m-1)|a|)(|b-1|+(n-1)|b|).
\end{eqnarray}%
For any other combinations of $\mathcal{Y}$ from $\{r_{A},c_{A},r_{B},c_{B}\}$,
following the same procedure above, we arrive at the final result
(\ref{maintheorem}). \hfill \rule{1ex}{1ex}

\subsubsection{Special cases reducing to other necessary criteria}

We show now that the Theorem actually encompasses two previous strong
computational criteria for separability.

\paragraph{The reduction criterion}

For the case of $a=0$ and $b=1$, we have
\begin{equation}
||\widetilde{\rho _{AB}}^{\mathcal{T}_{\mathcal{Y}}}||\leq \left\{
\begin{array}{l}
(n-1),\text{ \ \ \ }r_{B}\text{,}\,c_{B}\in \mathcal{Y}\text{\ \ or }r_{B}
\text{,}\,c_{B}\notin \mathcal{Y}\text{,\ } \\
(n-1)^{\frac{1}{2}},\text{ \ }r_{B}\in \mathcal{Y}\text{ }or\text{ }c_{B}\in
\mathcal{Y}\text{.}%
\end{array}%
\right.
\end{equation}%
When $r_{B}$, $c_{B}\notin \mathcal{Y}$, we have further
\begin{equation}
||\rho _{A}\otimes \mathbb{I}_{n}-\rho _{AB}||\leq n-1.  \label{r1}
\end{equation}%
For the case of $a=1$ and $b=0$, one obtains similarly
\begin{equation}
||\mathbb{I}_{m}\otimes \rho _{B}-\rho _{AB}||\leq m-1.  \label{r2}
\end{equation}%
Eqs.~(\ref{r1}) and (\ref{r2}) are equivalent to the reduction criterion,
since for separable states positivity of Eq.~(\ref{reduction}) means that the
trace norm is the sum of the eigenvalues, that is, the singular values,
due to the Hermitian property of $\rho _{A}\otimes \mathbb{I}_{n}-\rho _{AB}$ and
$\mathbb{I}_{m}\otimes \rho _{B}-\rho _{AB}$ \cite{hornt}.

\paragraph{The GPT criterion}

\label{2.3.2.b}

The \emph{GPT} criterion derived in \cite{chenPLA02} says that for any
\emph{GPT} operations, the trace norm for the realigned matrix is not greater than
$1$. Violation of that inequality means existence of entanglement. The \emph{GPT}
criterion includes the realignment criterion and the $PPT$ criterion as
special cases. Now the \emph{GPT} criterion is just one of the special cases
of our generalized reduced criterion: $a=0$ and $b=0$. In this case we have
\begin{equation}
||\rho _{AB}^{\mathcal{T}_{\mathcal{Y}}}||\leq 1,~~~\forall \mathcal{Y}\subset
\{r_{A},c_{A},r_{B},c_{B}\}.
\end{equation}
This is exactly the \emph{GPT} criterion.

\subsubsection{Invariance of our generalized reduction
criterion under local unitary transformations}

The trace norm $||\widetilde{\rho _{AB}}^{\mathcal{T}_{\mathcal{Y}}}||$ is
invariant under local unitary transformations. To see this, let $W_{A}$ (resp. $W_{B}$)
be unitary transformations on the subsystem $A$ (resp. $B$).
Under the local unitary transformation $W_{A}\otimes W_{B}$,
$\widetilde{\rho _{AB}}$ is mapped to $\widetilde{\rho _{AB}^{^{\prime }}}
=(W_{A}\otimes W_{B})\widetilde{\rho _{AB}}
(W_{A}^{\dagger }\otimes W_{B}^{\dagger })$. If for any \emph{GPT}
operations, the local unitary transformation only contributes some unitary factors to
$\widetilde{\rho _{AB}}^{\mathcal{T}_{\mathcal{Y}}}$, we would certainly have
$||\widetilde{\rho _{AB}}^{\mathcal{T}_{\mathcal{Y}}}||=||\widetilde{\rho
_{AB}^{^{\prime }}}^{\mathcal{T}_{\mathcal{Y}}}||$, due to the unitary
invariance of the trace norm. In fact, $\widetilde{\rho _{AB}}$
has the decomposition $\widetilde{\rho _{AB}}
=\sum_{i=1}^{q}\alpha _{i}\otimes \beta _{i}$, where $q=\min (m^{2},n^{2})$,
in terms of the procedure (\ref{trans}) to (\ref{sum}). Without loss of
generality, we consider a row transposition on the subsystem $A$.
Setting $\gamma _{i}=(W_{A}\otimes W_{B})(\alpha _{i}\otimes \beta
_{i})(W_{A}^{\dagger }\otimes W_{B}^{\dagger }),$ we have
\begin{eqnarray}
&&\gamma _{i}^{\mathcal{T}_{\{r_{A}\}}}=\mathcal{V}_{vec}(W_{A}\alpha _{i}W_{A}^{\dagger
})^{t}\otimes W_{B}\beta _{i}W_{B}^{\dagger }  \notag \\
&=&(\mathcal{V}_{vec}(\alpha _{i}))^{t}(W_{A}{}^{\dag }\otimes W_{A}^{t})\otimes
W_{B}\beta _{i}W_{B}^{\dagger }  \notag \\
&=&W_{B}\big((\mathcal{V}_{vec}(\alpha _{i}))^{t}\otimes \beta _{i}\big)(W_{A}{}^{\dag
}\otimes W_{A}^{t}\otimes W_{B}^{\dagger })
\end{eqnarray}
Summing over all the components $\gamma _{i}^{\mathcal{T}_{\{r_{A}\}}}$,
$i=1,2,...,q$, we have $\widetilde{\rho _{AB}^{^{\prime }}}^{\mathcal{T}_{\{r_{A}\}}}
=W_{B}\widetilde{\rho _{AB}}^{\mathcal{T}_{\{r_{A}\}}}(W_{A}{}^{\dag }\otimes
W_{A}^{t}\otimes W_{B}^{\dagger })$. Therefore
$\widetilde{\rho _{AB}^{^{\prime }}}^{\mathcal{T}_{\{r_{A}\}}}$
and $\widetilde{\rho _{AB}}^{\mathcal{T}_{\{r_{A}\}}}$ are equivalent under the
unitary factors $W_{B}$ and $(W_{A}{}^{\dag }\otimes W_{A}^{t}\otimes
W_{B}^{\dagger })$, which keep the trace norm invariant.

The same procedure can be used to perform column transposition, partial
transposition of some subsystems, and any combinations of these \emph{GPT}
operations, to show that the trace norm is an invariant under
local unitary transformation.

\section{Examples}\label{sec3}

In this section, we consider two examples to illustrate some
special characters of the criterion compared with the previously known reduction
criterion and the \emph{GPT} criterion.

\vspace*{12pt} \noindent \emph{Example 1:} $3\times 3$ Werner state.

\begin{figure}[tbp]
\begin{center}
\resizebox{8cm}{!}{\includegraphics{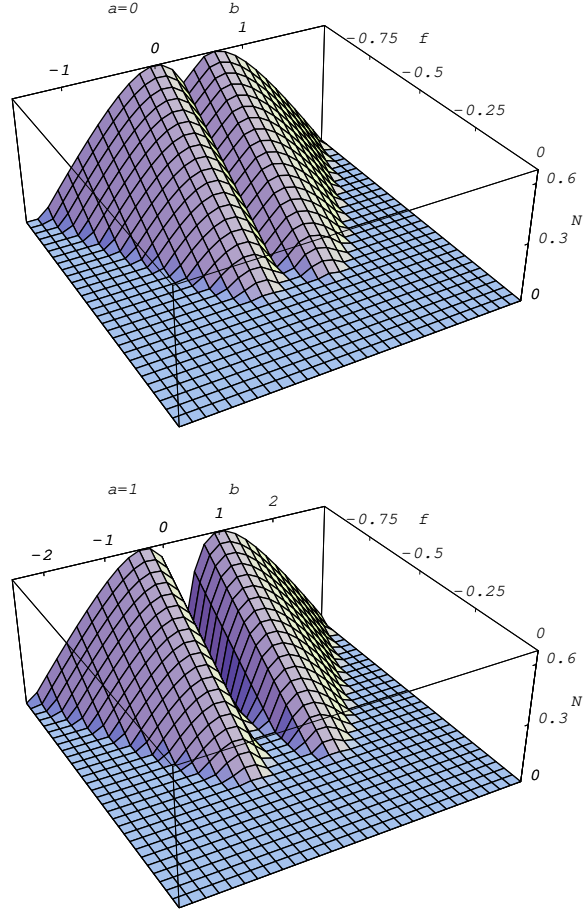}}
\end{center}
\caption{Depiction of $N=\max \{||\widetilde{\protect\rho _{AB}}^{\mathcal{T}%
_{\{c_{A},r_{B}\}}}||-h_{a}h_{b},0\}$ for a $3\times 3$ Werner state as a
function of $b$ and $f$ when $a=0$ (the top figure) and $a=1$ (the bottom
figure), respectively.}
\label{fig1}
\end{figure}

Consider the family of $d$-dimensional Werner states \cite{werner89},
\begin{equation}
W_{d}\equiv \frac{1}{d^{3}-d}\big(\left( d-f\right) \mathbb{I}_{d^{2}}+(df-1)V\big),
\label{werner}
\end{equation}
where $-1\leq f\leq 1$, $V(\alpha \otimes \beta )=\beta \otimes \alpha $,
the operator $V$ has the representation $V=\sum_{i,j=0}^{d-1}|ij\rangle \langle ji|$,
and $W_d$ is non-separable if and only if $-1\leq f<0$. As
is well known, the entanglement in a $2\times 2$ Werner state can be detected
completely using \emph{PPT}, reduction and realignment criteria. But for
higher dimensions the reduction criterion fails while \emph{PPT} succeeds.
The realignment criterion can
recognize entanglement for $-1\leq f<\frac{2}{d}-1$ \cite{ru02}.
Here for simplicity we consider the $3\times 3$
Werner state given by (\ref{werner}) and take $a,b\in\Rb$. We plot $N=\max \{||%
\widetilde{\rho _{AB}}^{\mathcal{T}_{\{c_{A},r_{B}\}}}||-h_{a}h_{b},0\}$ as
a function of $b$ and $f$ for the cases of $a=0$ and $a=1$ respectively,
Fig.\ref{fig1}.

For the case $a=0$, we see that $N>0$ for all $-1\leq f<-\frac{1}{3}$ and $b=0$
or $\frac{2}{3}$. From the Theorem, we have $h_{a}h_{b}=1$ for $b=0$ or $%
\frac{2}{3}$ when $a=0$. $\widetilde{\rho _{AB}}$ is in fact the same for $%
b=0$ or $\frac{2}{3}$ and we obtain $N=\max \{\frac{|1-3f|-2}{3},0\}$ by
direct computation. For that case $a=1$, we still have $N>0$ for all
$-1\leq f<-\frac{1}{3}$ but with $b=-\frac{1}{3}$ or $1$. Also we see that $h_{a}h_{b}=2$
for $b=-\frac{1}{3}$ or $1$ when $a=1$. $\widetilde{\rho _{AB}}$ is
the same for both $b=-\frac{1}{3}$ and $1$. In this case we have again
$N=\max \{\frac{|1-3f|-2}{3},0\}$ by direct verification.

\vspace*{12pt} \noindent \emph{Example 2:} Horodecki $3\times 3$ bound
entangled state

\begin{figure}[tbp]
\begin{center}
\resizebox{8cm}{!}{\includegraphics{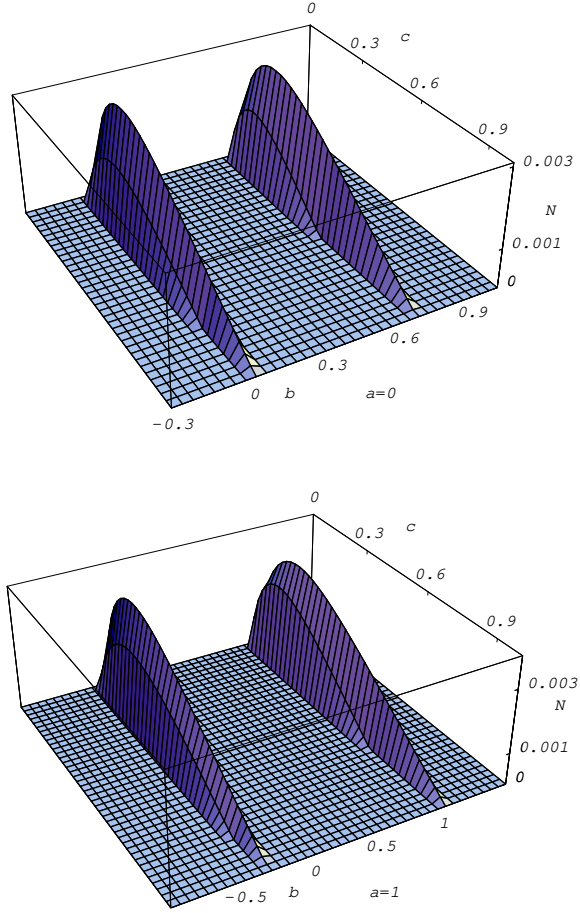}}
\end{center}
\caption{Depiction of $N=\max \{||\widetilde{\protect\rho _{AB}}^{\mathcal{T}%
_{\{c_{A},r_{B}\}}}||-h_{a}h_{b},0\}$ for a Horodecki $3\times 3$ bound
entangled state as a function of $b$ and $c$ when $a=0$ (the top figure) and
$a=1$ (the bottom figure), respectively.}
\label{fig2}
\end{figure}

Horodecki gives a very interesting weakly entangled state in \cite{hPLA97}
which cannot be detected by the $PPT$ criterion. The density matrix $\rho $
is real and symmetric:
\begin{equation}
\rho ={\frac{1}{8c+1}}\left[
\begin{array}{ccccccccc}
c & 0 & 0 & 0 & c & 0 & 0 & 0 & c \\
0 & c & 0 & 0 & 0 & 0 & 0 & 0 & 0 \\
0 & 0 & c & 0 & 0 & 0 & 0 & 0 & 0 \\
0 & 0 & 0 & c & 0 & 0 & 0 & 0 & 0 \\
c & 0 & 0 & 0 & c & 0 & 0 & 0 & c \\
0 & 0 & 0 & 0 & 0 & c & 0 & 0 & 0 \\
0 & 0 & 0 & 0 & 0 & 0 & {\frac{1+c}{2}} & 0 & {\frac{\sqrt{1-c^{2}}}{2}} \\
0 & 0 & 0 & 0 & 0 & 0 & 0 & c & 0 \\
c & 0 & 0 & 0 & c & 0 & {\frac{\sqrt{1-c^{2}}}{2}} & 0 & {\frac{1+c}{2}}%
\end{array}%
\right] ,
\end{equation}%
where $0<c<1$. The entanglement in this state is very difficult to detect
with previous operational criterion in general. In \cite{hPLA97}, Horodecki
showed that the \emph{range} criterion could recognize the entanglement. For
this state, the simple realignment criterion and its generalization: the
\emph{GPT} criterion can, surprisingly, detect the entanglement for all
permissable $c$, as shown in \cite{ChenQIC03}. Here, we will observe the
behavior in the language of the generalized reduction criterion
(for real values of $a$ and $b$, as in Example $1$).
The partial transposition $\widetilde{\rho _{AB}}^{\mathcal{T}%
_{\{r_{A},c_{A}\}}}$ using our Theorem fails to detect the entanglement by
direct calculation. Therefore we are only concerned with the case of $\widetilde{%
\rho _{AB}}^{\mathcal{T}_{\{c_{A},r_{B}\}}}$ which corresponds to the
realignment of $\widetilde{\rho _{AB}}$. For the case of $a=0$ and $a=1$,
respectively, we plot $N=\max \{||\widetilde{\rho _{AB}}^{\mathcal{T}%
_{\{c_{A},r_{B}\}}}||-h_{a}h_{b},0\}$ as a function of $b$ and $c$ in Fig.%
\ref{fig2}.
\noindent For the case of $a=0$, we see that $N>0$ for all $0<c<1$
and $b=0$ or $\frac{2}{3}$. In this case $h_{a}h_{b}=1$
for $b=0$ or $\frac{2}{3}$ when $a=0$. But $\widetilde{\rho _{AB}}$ has
different forms for $b=0$ and $\frac{2}{3}$. For the case $a=1$ we still have
$N>0$ for all $0<c<1$ but with $b=-\frac{1}{3}$ or $1$. Also we see that $h_{a}h_{b}=2$
and $\widetilde{\rho _{AB}}$ has quite different forms for $b=-\frac{1}{3}$ and $1$
when $a=1$. However, the function $N$ has the same value for the case
$a=0 $ when $b=0$ or $\frac{2}{3}$, or the case $a=1$ when $b=-\frac{1}{3}$
or $1$.

\vspace*{12pt}
The results of the above two examples are a little bit surprising compared
with the relationship between the \emph{PPT} and reduction criteria. As is
well known, the $PPT$ criterion is equivalent to the reduction criterion for a
$2\times n$ system$.$ That is to say, for $a=0$ and $b=1$, or $a=1$ and $b=0$
they give the same result for a $2\times n$ system. But in the case of the $%
3\times 3$ system, we have identical results for different $b$ other than
the value of $0,1$ if we apply the realignment operations to $\widetilde{\rho
_{AB}}$. This interesting phenomenon also occurs in higher dimensional cases.
So our generalized reduction includes all the results from
the original realignment or $GPT$\ criteria but recognizes entanglement in
a very subtle way which is quite different from other criteria.

\section{Conclusion}

\label{sec4} Summarizing, we have introduced a computational
criterion, which we call the \textquotedblleft generalized reduction
criterion", providing a necessary condition
for separability of bipartite quantum systems in arbitrary
dimensions . This criterion combines many virtues of the reduction criterion
and the \emph{GPT} criterion. It gives a unified framework for the
two criteria and provides a powerful necessary condition for
separability using just simple matrix operations. Some interesting characters
of this criterion are showed by two typical examples. The theorem can be
straightforwardly generalized to the multipartite case
by introducing more free parameters like $a$ and $b$ in the theorem.

We expect that this construction not only expands
theoretically our sight in detecting
entanglement of a general quantum state, but also sheds some
light on possible ways to the final solution of the separability
problem. We conjecture that any future stronger operational separability
test should in principle include as special cases the simple \emph{GPT}
criterion, in particular the \emph{PPT} criterion which is necessary and
sufficient one for $2\times 2$ and $2\times 3$ system. This paper is an
attempt following this way, though we have not yet found an example of a bound
entangled state which can be detected by this criterion but not by
the \emph{GPT} criterion.

\section*{Acknowledgments}

The work is supported by SFB611. K. Chen gratefully acknowledges
the hospitality of the Institute of Applied Mathematics of Bonn University. He
would also like to thank Prof. Guozhen Yang and Prof. Ling-An Wu for their
continuous encouragements. He has been partly supported by the China
Postdoctoral Science Foundation.

\end{document}